\documentclass[aps, prl, twocolumn, superscriptaddress, amsmath, showpacs, tightenlines, footinbib, longbibliography]{revtex4-1}
\usepackage[colorlinks, breaklinks, citecolor=blue, linkcolor=blue,  urlcolor={blue}]{hyperref}
\usepackage{breakurl}
\usepackage{amsfonts}
\usepackage{amsmath}
\usepackage{amssymb}
\usepackage{mathrsfs}
\usepackage{epsfig}
\usepackage{graphicx}
\usepackage{bm}
\usepackage{subfigure}

\usepackage[dvipsnames]{xcolor}

\usepackage{hyperref}
\begin{document}
	\title{Two-Polariton Blockade via Ultrastrong Light-Matter Coupling
	}
	
	\author{Ting-Ting Ma}
	\affiliation{Key Laboratory of Low-Dimensional Quantum Structures and Quantum Control of Ministry of Education, Department of Physics and Synergetic Innovation Center for Quantum  Effects and Applications, Hunan Normal University, Changsha 410081, China}
	
	\author{Jian Tang}
	\affiliation{Key Laboratory of Low-Dimensional Quantum Structures and Quantum Control of Ministry of Education, Department of Physics and Synergetic Innovation Center for Quantum Effects and Applications, Hunan Normal University, Changsha 410081, China}
	
	\author{Yun-Lan Zuo}
	\affiliation{School of Physics and Chemistry, Hunan First Normal University, Changsha 410205, China}
	
	\author{Ran Huang}
	\email{ran.huang@riken.jp}
	\affiliation{Quantum Information Physics Theory Research Team, Center for Quantum Computing (RQC), RIKEN, Wakoshi, Saitama 351-0198, Japan}
	
	\author{Adam Miranowicz}
	\affiliation{Institute of Spintronics and Quantum Information, Faculty of Physics, Adam Mickiewicz University, 61-614 Pozna{\'{n}}, Poland}
	
	\author{Franco Nori}
	\affiliation{Quantum Information Physics Theory Research Team, Center for Quantum Computing (RQC), RIKEN, Wakoshi, Saitama 351-0198, Japan}
	\affiliation{Department of Physics, University of Michigan, Ann Arbor, Michigan 48109-1040, USA}
	
	\author{Hui Jing}
    \email{jinghui73@foxmail.com}
	\affiliation{Key Laboratory of Low-Dimensional Quantum Structures and Quantum Control of Ministry of Education, Department of Physics and Synergetic Innovation Center for Quantum Effects and Applications, Hunan Normal University, Changsha 410081, China}
	\affiliation{Institute for Quantum Science and Technology, College of Science, National University of Defense Technology, Changsha 410073, People’s Republic of China}
	
	\begin{abstract}
		We demonstrate that a two-polariton blockade (2PB) can occur under resonant single-polariton driving in an atom-cavity system operating in the ultrastrong coupling (USC) regime---a phenomenon qualitatively distinct from, and unattainable in, both the strong and weak coupling regimes. In the USC regime, where the ratio of the atom-cavity coupling strength to the cavity resonance frequency exceeds 0.1, hybrid light-matter quasiparticles known as polaritons emerge. By employing modified second- and third-order correlation functions appropriate for the USC regime, we predict the emergence of 2PB, characterized by pronounced two-polariton bunching accompanied by suppressed three-polariton coincidences. This Letter introduces a novel route to achieving 2PB, with promising implications for the realization of multiparticle quantum light sources in the USC regime.
		
	\end{abstract}
	
	\maketitle
	
	Ultrastrong coupling (USC) between light and matter refers to a regime where the coupling strength becomes comparable to the cavity mode frequency~\cite{frisk2019ultrastrong, RevModPhys.91.025005, Adv.Quantum.Technol.3.7, mischok2024breaking, PhysRevLett.108.120501, anappara2009signatures, ashhab2010qubit, 2010Circuit}. In this USC regime, the rotating wave approximation is no longer valid, meaning that the counter-rotating terms (CRTs) are non-negligible. Due to its highly efficient interactions, USC has been extensively studied in various fields, including quantum metrology~\cite{garbe2020critical, PhysRevA.100.053825, gietka2022understanding}, quantum plasmonics~\cite{tame2013quantum, benz2016single, saez2023can}, polariton-enhanced superconductivity~\cite{schlawin2019cavity}, metamaterials~\cite{scalari2012ultrastrong, bayer2017terahertz}, and quantum thermodynamics~\cite{PhysRevE.98.012131}. USC has been realized in various experimental systems, such as intersubband polaritons~\cite{anappara2009signatures, gunter2009sub, todorov2010ultrastrong, delteil2012charge, geiser2012ultrastrong}, superconducting quantum circuits~\cite{PhysRevLett.105.237001, forn2017ultrastrong, yoshihara2017superconducting, yoshihara2017characteristic, yoshihara2018inversion}, Landau polaritons~\cite{scalari2012ultrastrong, PhysRevB.101.075301}, organic molecules~\cite{schwartz2011reversible, gubbin2014low, mazzeo2014ultrastrong, barachati2018tunable}, optomechanics~\cite{george2016multiple, benz2016single}, and magnons~\cite{makihara2021ultrastrong, PhysRevLett.123.117204}. Due to the extensive application of circuit quantum electrodynamics (cQED) systems~\cite{2010Vacuum, 2017Generation, 2019Ultrastrong, PhysRevLett.107.053602, PhysRevA.97.013851, PhysRevB.60.15398, mooij1999josephson, PhysRevLett.120.093601, PhysRevLett.120.093602, 10.1063/1.5089550, annurev-conmatphys-031119-050605, GU20171}, achieving USC between light and matter in experiments has become more feasible, and the study of quantum effects in USC systems has attracted widespread attention~\cite{QIN20241, Nat.Commun.12.6206, s41467-020-16524-x, PhysRevLett.123.247701, wang2023magnetically, PhysRevLett.126.153603, yu2023strong, PhysRevLett.126.023602, di2019resolution}.
	USC gives rise to various quantum phenomena that are absent in weak and strong coupling regimes. These include ultrafast quantum gate operations~\cite{PhysRevLett.108.120501, PhysRevA.96.063820}, enhanced coherence times and quantum operation fidelities~\cite{PhysRevLett.107.190402, PhysRevLett.107.190402}, as well as persistent quantum memory \cite{PhysRevA.97.033823, kyaw2015scalable, PurcellEffect}. However, while USC enhances many quantum phenomena, it can also weaken quantum correlations in single-quasiparticle control. Notably, bunching occurs under a certain resonant single-polariton excitation~\cite{PhysRevLett.109.193602, PhysRevA.94.033827}, which can make USC less effective than strong coupling (SC) for single-quasiparticle manipulation. This effect arises because, although USC enhances level anharmonicity, the counter-rotating wave terms simultaneously introduce additional multipolariton channels. When appropriate driving fields are applied, they may excite multiple transition pathways at once, thereby undermining the nonlinearity typically required for single-polariton blockade (1PB).
	
	One well-known nonlinear effect in SC systems is polariton blockade, in which strong coupling induces energy level splitting, increasing the energy gap between multipolariton states and thereby suppressing the generation of additional polaritons~\cite{PhysRevB.73.193306, PhysRevLett.125.197402}. 1PB is analogous to photon blockade: when a single polariton exists in the system, it prevents the formation of a second polariton~\cite{birnbaum2005photon, huang2022exceptional, PhysRevLett.121.153601, PhysRevA.105.053718, PhysRevA.101.063838, birnbaum2005photon}. This effect is useful for single-quasiparticle sources and quantum information processing, providing a means for efficient single-quasiparticle manipulation~\cite{PhysRevLett.83.4204, o2009photonic, reiserer2015cavity}. Multi-polariton blockade (MPB) is an extension of 1PB, focusing on the interactions and blockade effects among multipolaritons~\cite{hartmann2006strongly}. Achieving MPB requires stronger nonlinear interactions~\cite{greentree2006quantum, PhysRevA.87.023809}. In the SC regime, MPB has been extensively studied~\cite{PhysRevA.87.023809, PhysRevLett.118.133604, wang2020photon, kowalewska2019two}, demonstrating potential for generating entangled quasiparticle pairs and multiquasiparticle sources~\cite{PhysRevA.93.013808, PhysRevLett.101.203602, PhysRevA.102.053710, liu2023deterministic, PhysRevA.98.043858}. However, to our knowledge, MPB in USC systems remains largely unexplored, and its underlying physical mechanisms remain poorly understood. This raises a critical question: do novel quantum effects emerge in USC systems when 1PB vanishes~\cite{PhysRevLett.109.193602}, and are other quantum effects enhanced?
	
	Here, we propose a counterintuitive phenomenon: in the USC regime, two-polariton blockade (2PB) may occur under resonant single-polariton excitation. Our findings reveal fundamental differences between the nature of polariton blockade in the USC and SC regimes. In the SC regime, only 1PB occurs under resonant single-polariton excitation. However, in the USC regime, by calculating the equal-time second- and third-order correlation functions, we demonstrate that within the appropriate parameter range, not only does 1PB occur, but notably, 2PB can also emerge at a specific resonant single-polariton excitation---a phenomenon qualitatively distinct and inaccessible in both strong and weak coupling regimes. To rigorously establish 2PB, we apply criteria based on polariton antibunching, demonstrating that genuine 2PB is characterized by three-polariton antibunching accompanied by two-polariton bunching. Further analysis of transition rates between different dressed states and the population distributions confirms the presence of this unique 2PB under resonant single-polariton excitation within the USC regime. The 2PB effect unveils a practical mechanism for multiphoton sources~\cite{PhysRevLett.118.133604, PhysRevA.87.023809, PhysRevA.102.053710, carusotto2013quantum}. Excitation of the two polaritons in this atom-cavity system blocks the absorption of subsequent photons because of the anharmonicity of the eigenenergy level structure. The identified cascade decay inherently generates polariton pairs. Therefore, such system with 2PB effect converts a coherent light stream into a bunched photon pair emission, which is important for realizing two-photon sources. For higher-order excitations, three-polariton and multipolariton blockade effects hold the potential for realizing multiphoton light sources ~\cite{2015Enhancement, liu2023deterministic, RevModPhys.93.025005, 2010Circuit}.
	
\begin{figure}[t!]
		\centering
		\includegraphics[width=0.52\textwidth]{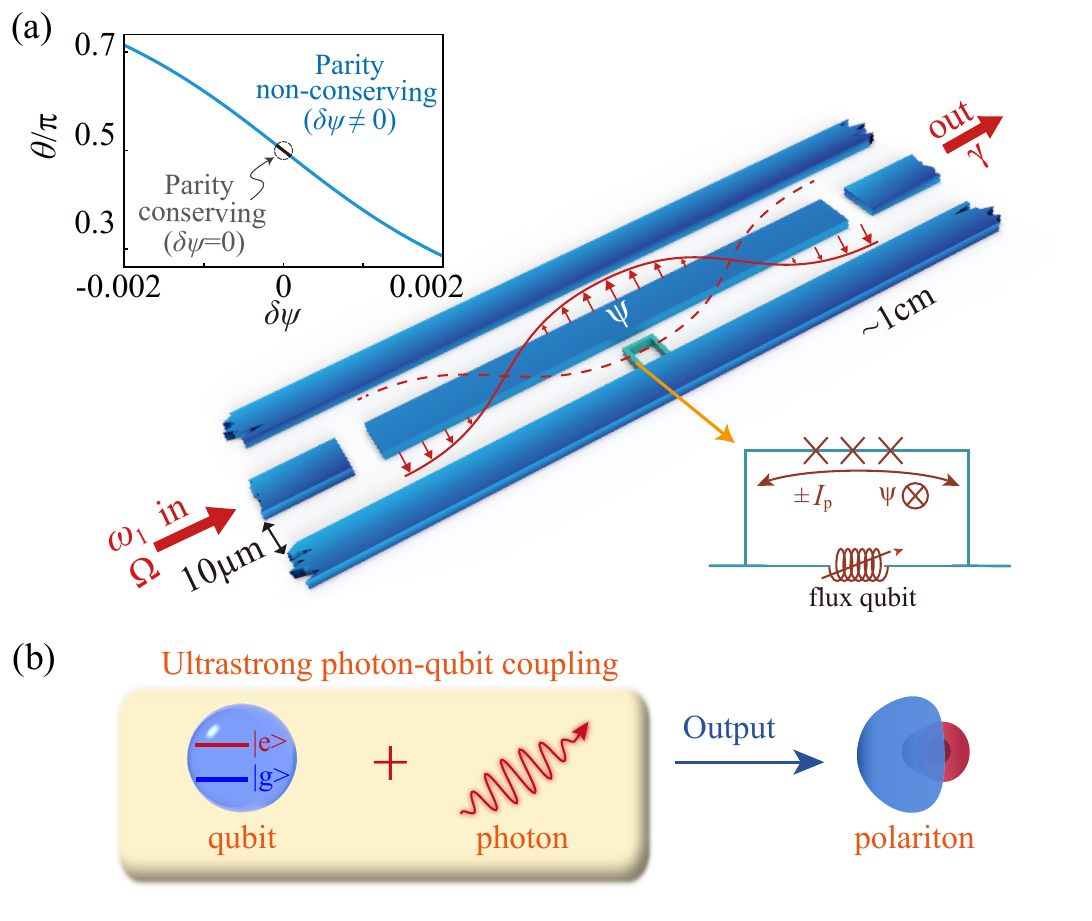}
		\caption{(a) Schematic of the cQED system under study. A transmission-line resonator is coupled to a flux qubit, which is tunable via an external magnetic flux bias $\psi$. Here, $\pm I_p$ denote the clockwise and counterclockwise persistent currents circulating in the qubit loop. Parity is conserved when $\theta=m \pi/2$ ($\delta \psi=0$). For $\theta=0.3\pi$, the parity is non-conserved. The cavity mode is driven externally (in red) with frequency $\omega_l$ and amplitude $\Omega$ via the input capacitor, while coherent transmission is measured at the output capacitor. (b) Illustration of polariton formation via ultrastrong coupling between the qubit and the photon.}
		\label{Fig1}
	\end{figure}

	\emph{Model}---As shown in Fig.~\ref{Fig1}, we consider a cQED system with general linear coupling between a flux qubit and a transmission-line resonator. This setup effectively models a two-level atom interacting with a single cavity mode for demonstrating quantum effects. The cavity mode is driven by a weak coherent laser field. The total Hamiltonian reads
	\begin{align}
		&H=H_0+H_{\mathrm{drive}},\nonumber\\
		&H_0=\omega_c a^{\dagger}a+\omega_g \sigma^{+}\sigma^{-}+g(a+a^{\dagger})(\cos \theta \sigma_z-\sin \theta \sigma_x),\nonumber\\
		&H_{\mathrm{drive}}=\Omega \cos(\omega_l t)(a+a^{\dagger}),
		\label{Hami}
	\end{align}
	where $H_0$ describes the system energy, including both the atom and the cavity mode. The operators $a$ and $a^{\dagger}$ are the annihilation and creation operators of the cavity mode, respectively. The Pauli operators are defined as $\sigma^{+}=\left\vert e \right\rangle \left\langle g \right\vert$, $\sigma_x=\sigma^{+}+\sigma^{-}$, and $\sigma_z=\sigma^{+}\sigma^{-}-\sigma^{-}\sigma^{+}$. Here, $\omega_c$ is the cavity resonance frequency, $\omega_g$ is the atomic transition frequency, and $g$ is the coupling strength. In USC systems, the normalized coupling strength $g/\max[\omega_c,\omega_g]$ exceeds $0.1$~\cite{frisk2019ultrastrong, RevModPhys.91.025005}. The term $H_{\rm drive}$ represents the coherent driving of the
	cavity mode, with $\Omega$ and $\omega_l$ denoting the drive amplitude and drive frequency, respectively. The atom and cavity interaction in $H_0$ describes a general linear coupling~\cite{PhysRevLett.129.066801, PhysRevA.80.032109, deppe2008two, yoshihara2017superconducting}. $\theta$ determines the system's parity~\cite{PhysRevLett.109.193602, xu2023anti, 2010Circuit} and is influenced by the bias flux through the qubit loop. It is given by $\cos\theta=2I_p \delta \psi/\sqrt{\Delta^2+(2I_p\delta \psi)^2}$, where $\pm I_p$ are the clockwise and anticlockwise persistent currents along the qubit loop, $\Delta$ is the sweet-spot frequency ($\delta \psi=0$), while the coupling phase $\theta$ is tuned by varying the external flux bias. When $\theta=m \pi/2$, the excitation-number parity is conserved, this symmetry breaks down when $\theta \ne m \pi/2$~\cite{2010Circuit, jiang2024dual}. In the latter case, more complex transitions can arise in USC systems. These transitions, distinct from those in weaker coupling regimes, permit transitions between arbitrary states and lead to exotic effects in polariton statistics.
	
	In summary, the weakening of 1PB in the USC regime primarily arises under conditions of parity nonconservation ($\theta\ne m \pi/2$). Although numerous studies have investigated polariton blockade in USC systems~\cite{PhysRevLett.109.193602, PhysRevA.94.033827, QIN20241}, high-order correlation effects remain largely unexplored. Here, we focus on how multipolariton processes in USC systems differ from those in SC systems by analyzing high-order correlation functions.

	To accurately model dissipative processes in the USC regime, we diagonalize the total system Hamiltonian $H_0$ to obtain the dressed states satisfying $H_0 \sum_j \left\vert \psi_j \right\rangle= \sum_j \hbar \omega_j \left\vert \psi_j \right\rangle (\hbar=1)$. The eigenstates of the entire system are denoted by $\left\vert \psi_{j} \right\rangle$. By expanding the dissipative terms in the eigenbasis and following the standard procedure for deriving the master equation~\cite{2012Spontaneous, johansson2012qutip, johansson2013qutip2,  gonzalez2024light, Li2022pulselevelnoisy, lambert2024qutip}, we have the master equation as $\dot\rho(t) = i [\rho(t), H] + \mathcal{L}_{a}\rho(t)+\mathcal{L}_{\sigma^{-}}\rho(t)$. Here, $\mathcal{L}_{c}\rho(t) = \sum_{j,k>j}\Gamma^{j k}_{c} \mathcal{D}[|\psi_{j} \rangle \langle \psi_{k}|]\rho(t)(c=a,\sigma^{-})$ is the Liouvillian superoperator denoting the losses of the system, and the dissipator is defined as $\mathcal{D}[\mathcal{O}]\rho = \frac{1}{2} (2 \mathcal{O}\rho\mathcal{O}^{\dagger}-\rho \mathcal{O}^{\dagger} \mathcal{O} - \mathcal{O}^{\dagger} \mathcal{O}\rho)$. We consider the voltages of the resonator and output line couple via a capacitance, which is the same as in Refs.~\cite{9zl7-31f3, PhysRevLett.110.163601, PhysRevLett.109.193602, PhysRevLett.117.043601}. We assume the spectral density $d_{c}(\Delta_{k j})$ is constant and system-bath coupling strength $\alpha^{2}_{c}(\Delta_{k j})$ is proportional to the transition frequency difference $\Delta_{kj}$~\cite{PhysRevLett.110.163601, PhysRevLett.109.193602, PhysRevLett.117.043601}. Therefore, the relaxation coefficients $\Gamma^{j k}_{c} = 2\pi d_{c}(\Delta_{k j}) \alpha^{2}_{c}(\Delta_{k j})| C^{(c))}_{j k}|^2$, can be simplified to $\Gamma^{j k}_{c} = \gamma_{c}\frac{\Delta_{k j}}{\omega_0}  |C^{(c)}_{j k}|^2$ with $C^{(c)}_{j k} = -i \langle \psi_{j} |(c - c^{\dagger})| \psi_{k} \rangle(c=a,\sigma^{-})$, where $\gamma_c$ is the standard decay rate in the weak-coupling limit. 
	
	\begin{figure}[tpb]
		\centering
		\includegraphics[width=0.49\textwidth]{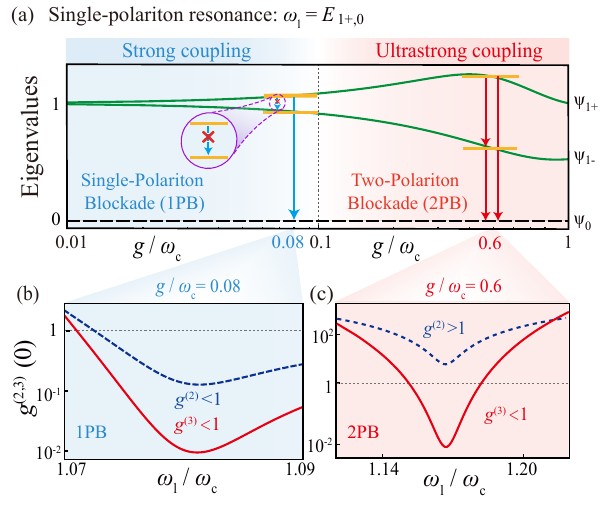}
		\caption{Polariton statistics for the SC and USC systems. (a) Energy spectrum as a function of coupling strength $g$. Illustrations of energy-level transitions for $g/\omega_c=0.08$  and $g/\omega_c=0.6$. (b), (c) The second-order (blue dashed curves) and third-order (red solid curves) equal-time correlation functions for $g/\omega_c =0.08$ and $g/\omega_c=0.6$, respectively. The dissipation parameters are $\gamma_c=\gamma_{\sigma^{-}}=\gamma=10^{-2}\omega_c$, the drive amplitude is $\Omega=10^{-1}\gamma $, and the coupling phase $\theta=0.3\pi$.}
		\label{Fig2}
	\end{figure}
	\emph{High-order correlations}---In the USC regime, the input-output relation is modified as $a_{\mathrm{out}}(t)=a_{\mathrm{in}}(t)-i\Upsilon \dot{X}^{+}$~\cite{PhysRevLett.109.193602}, where $\Upsilon$ denotes the waveguide coupling coefficient. We define the $n$th-order correlation function of the output field as
	\begin{equation}
		g^{(n)}(0)=\left\langle \dot{X}^{-n}\dot{X}^{+n}\right\rangle\left\langle \dot{X}^{-}\dot{X}^{+}\right\rangle^{-n}.
		\label{gn}
	\end{equation}
	Here, $\dot{X}^{+}$ can be expanded in the eigenbasis as
	\begin{equation}
		\dot{X}^{+}=-i \sum_{j,k>j} \Delta_{kj} X_{jk} \left\vert \psi_j \right\rangle \left\langle \psi_k \right\vert,
	\end{equation}
	with $X_{jk}=\langle \psi_j | X | \psi_k\rangle$. This operator captures not only photonic properties but also atomic nonlinearities, representing a polaritonic annihilation operator. 
	\begin{figure*}[tpb]
		\centering
		\includegraphics[width=0.95\textwidth]{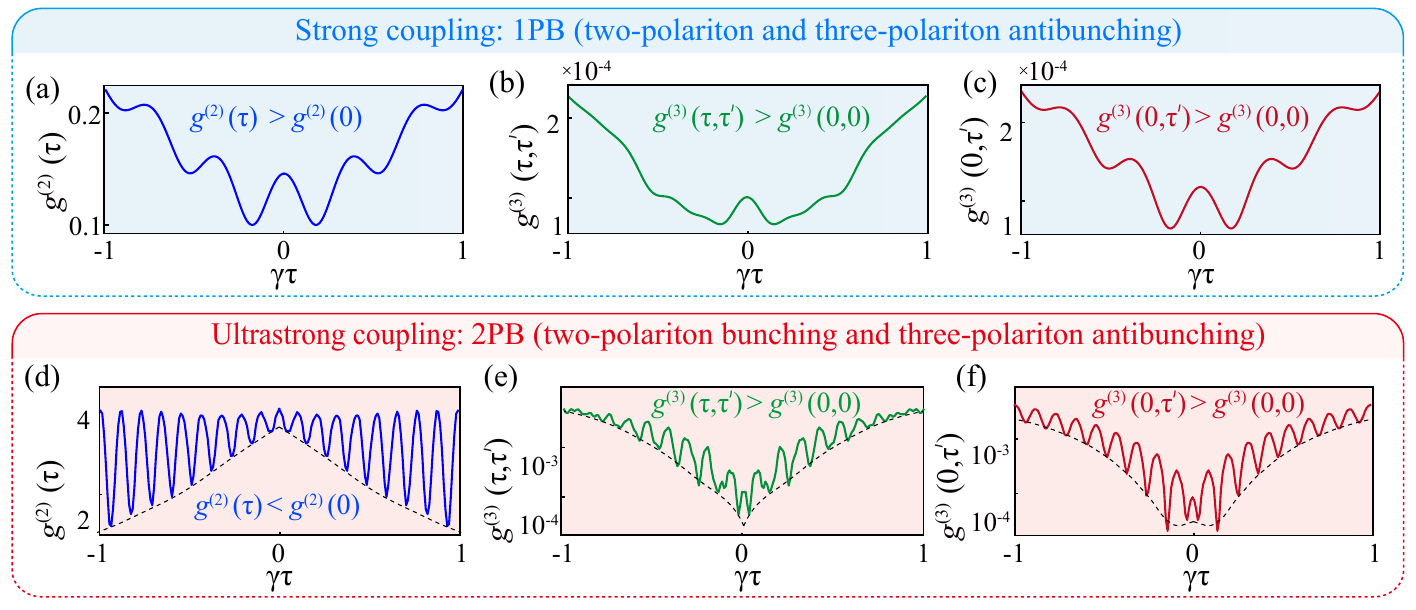}
		\caption{Delay-time correlation functions in the SC (a)–(c) and USC (d)–(f) regimes: (a), (d) $g^{(2)}(\tau)$ (blue solid curves), (b), (e) $g^{(3)}(\tau, \tau')$ (green solid curves), and (c), (f) $g^{(3)}(0, \tau')$ (red solid curves). The black dashed curves in (d)-(f) are the envelopes of the valleys of the correlation functions. Other parameters are the same as in Fig.~\ref{Fig2}. These results reveal three-polariton antibunching and two-polariton bunching, which are essential for true two-polariton blockade.}
		\label{Fig4}
	\end{figure*}
	
	Figure~\ref{Fig2}(a) illustrates the evolution of the energy spectrum with coupling strength $g$. In the SC regime ($g=0.08\omega_c$), resonant driving excites the system to $\left\vert \psi_{1+} \right\rangle$, which decays predominantly via $\left\vert \psi_{1+} \right\rangle \to \left\vert \psi_{0} \right\rangle$, while the transition $\left\vert \psi_{1+} \right\rangle \to \left\vert \psi_{1-} \right\rangle$ remains forbidden. In contrast, in the USC regime ($g=0.6\omega_c$, $\theta=0.3\pi$), the system undergoes a cascade decay via $\left\vert \psi_{1+} \right\rangle \to \left\vert \psi_{1-} \right\rangle \to \left\vert \psi_{0} \right\rangle$, opening a two-polariton emission channel.
	For a more comprehensive analysis of two-polariton processes in the USC regime, we calculate the second-order and third-order equal-time correlation functions. As depicted in Figs.~\ref{Fig2}(b) and ~\ref{Fig2} (c), $g^{(2)}(0)$ and $g^{(3)}(0)$ are presented, respectively, as a function of $\omega_l$. The qubit resonant frequency is the same as the frequency of the cavity mode, i.e., $\omega_g=\omega_c$. The parameters used are $\gamma_a=\gamma_{\sigma^{-}}=\gamma=10^{-2}\omega_c$, and $\Omega=10^{-1}\gamma$ for both the SC and USC regimes. In the SC regime, at the frequency dip corresponding to resonant single-polariton excitation with $\omega_l=E_{1+,0}$, the equal-time correlation functions satisfy $g^{(2)}(0) < 1$ and $g^{(3)}(0) < 1$, which correspond to both two- and three-polariton antibunching, indicating the occurrence of 1PB. In contrast, in the USC regime, the conditions of two-polariton
	bunching ($g^{(2)}(0)> 1$) and simultaneously three-polariton antibunching ($g^{(3)}(0) < 1$) indicate the emergence of 2PB~\cite{Comment1, Comment2, PhysRevLett.118.133604, PhysRevLett.121.153601, kowalewska2019two, liu2023deterministic}. This result highlights that 2PB can occur even under a single-polariton resonance condition, which is an exotic quantum phenomenon unique to the USC regime.
	
	Moreover, we propose an additional criterion for identifying 2PB based on delay-time correlation functions. The $n$th-order delay-time correlation function is defined as~\cite{PhysRevLett.118.133604, kowalewska2019two}
	\begin{equation}
		g^{(n)}(\tau_1,\tau_2,\dots,\tau_{n-1})=\left\langle : \hat{n}(0)\hat{n}(\tau_1)\dots \hat{n}(\tau_{n-1}) :\right\rangle \left\langle \hat{n} \right\rangle^{-n},
		\label{g1}
	\end{equation}
	where $\hat{n}=\dot{X}^{-}(t)\dot{X}^{+}(t)$.
	Additional criteria for identifying 2PB can be derived from delay-time polariton correlations, which manifest as delay-time two-polariton bunching and three-polariton antibunching,
	respectively, as follows:
	\begin{align}
		&g^{(2)}(\tau)< g^{(2)}(0),\nonumber\\
		&g^{(3)}(\tau,\tau')>g^{(3)}(0,0),\quad g^{(3)}(0,\tau')>g^{(3)}(0,0),
		\label{Eq2}
	\end{align}
	which means that a genuine 2PB requires three-polariton antibunching accompanied by two-polariton bunching. 
	
	Figure~\ref{Fig4} depicts delay-time correlation functions in both the SC and USC regimes. As shown in Figs.~\ref{Fig4}(a)–\ref{Fig4}(c), in the SC regime, the third-order delay-time correlations $g^{(3)}(0,\tau')$ and $g^{(3)}(\tau,\tau')$ both exceed $g^{(3)}(0,0)$, while the second-order delay-time correlation $g^{(2)}(\tau)$ surpasses $ g^{(2)}(0) $. This behavior indicates the simultaneous occurrence of two- and three-polariton antibunching. In contrast, Figs.~\ref{Fig4}(d)–\ref{Fig4}(f) reveal that the delay-time correlations in the USC regime satisfy the criteria outlined in Eq.~(\ref{Eq2}), thereby providing evidence for the existence of genuine 2PB in this regime. 
	\begin{figure}[tpb]
		\centering
		\includegraphics[width=0.49\textwidth]{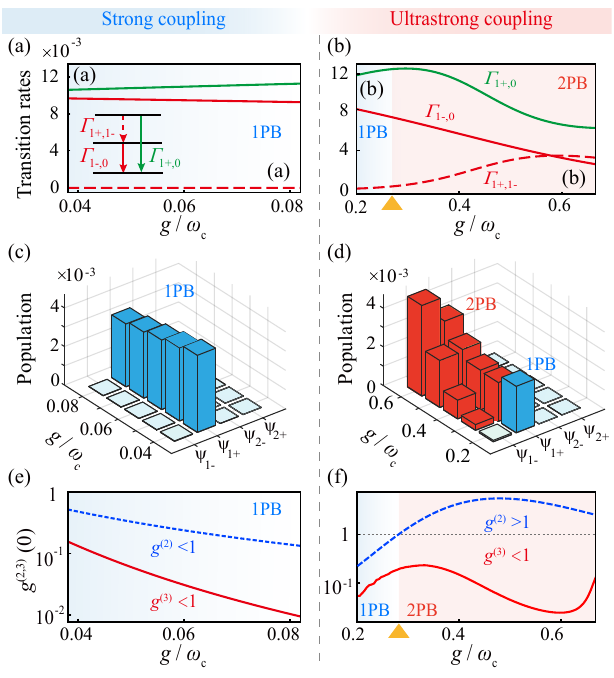}
		\caption{The physical mechanism of 2PB. (a), (b) Transition rates between the different dressed states $\Gamma_{1+,0}$ (the green solid line), $\Gamma_{1-,0}$ (the red solid line), $\Gamma_{1+,1-}$ (the red dashed line) versus the coupling strength under SC and USC regimes. (c), (d) The population of different dressed states under SC and USC regimes. $\theta = 0.3\pi$ and $\omega_l = E_{1+,0}$. (e) The equal-time correlation functions $g^{(2)}(0)$  and $g^{(3)}(0)$ as a function of the coupling strength $g$ in the SC regime. (f) Same as (e), but for the USC regime. The blue dashed and red solid curves represent the results of the second-order and third-order correlation functions, respectively. The yellow triangle point corresponds to $g/\omega_c=0.27$. Other parameters are the same as in Fig.~\ref{Fig2}.}
		\label{Fig3}
	\end{figure}
	
	\emph{Physical mechanism}---To explain the underlying physical mechanism behind this exotic behavior, we calculate the transition rates between the various dressed states.  Figure~\ref{Fig3}(a) shows the transition rates versus the coupling strength $g$. The green solid curves denote the rate for the transition $\left\vert \psi_{1+} \right\rangle \to \left\vert \psi_{0} \right\rangle$, the red solid curves denote the rate for the transition $\left\vert \psi_{1-} \right\rangle \to \left\vert \psi_{0} \right\rangle$, and the red dashed curves denote the rate for the transition $\left\vert \psi_{1+} \right\rangle \to \left\vert \psi_{1-} \right\rangle$. In the SC regime, $\Gamma_{1+,1-}$ remains zero, indicating that no transitions between the energy levels $\left\vert \psi_{1+} \right\rangle$ and  $\left\vert \psi_{1-} \right\rangle$. That is to say, in the SC regime, only single-polariton processes $\left\vert \psi_{1+} \right\rangle \to \left\vert \psi_{0} \right\rangle$ and $\left\vert \psi_{1-} \right\rangle \to \left\vert \psi_{0} \right\rangle$ are possible, while two-polariton cascade processes are forbidden. Entering into the USC regime, the transition rates between the dressed states undergo significant changes. When the coupling strength reaches approximately $0.27\omega_c$, the transition rate $\Gamma_{1+,1-}$ begins to increase significantly, indicating that the cascade process $\left\vert \psi_{1+} \right\rangle \to \left\vert \psi_{1-} \right\rangle \to \left\vert \psi_{0} \right\rangle$ becomes possible. This cascade mechanism promotes two-polariton bunching while suppressing three-polariton events—the defining signature of 2PB. We note that, such underlying mechanism can be further extended to other types of physical platforms, such as intersubband polaritons~\cite{anappara2009signatures, todorov2010ultrastrong, geiser2012ultrastrong}, organic molecules~\cite{schwartz2011reversible, george2016multiple}, Landau polaritons~\cite{scalari2012ultrastrong} and optomechanical systems~\cite{benz2016single}.

	To offer a cleaner explanation of how 2PB arises under resonant single-polariton excitation in the USC regime, we present the populations of different dressed states in both the SC and USC regimes in Figs.~\ref{Fig3}(c) and ~\ref{Fig3}(d). As shown in Fig.~\ref{Fig3}(c), in the SC regime, when the system is driven to state $\left\vert \psi_{1+} \right\rangle$, only this state $\left\vert \psi_{1+} \right\rangle$ has a population during the downward decay. That is because the transition from the state $\left\vert \psi_{1+} \right\rangle$ to the $\left\vert \psi_{1-} \right\rangle$ state is forbidden, permitting only the single-polariton processes. Figure~\ref{Fig3}(d) displays the populations of various dressed states in the USC regime. When the coupling strength is less than $0.27\omega_c$, the state $\left\vert\psi_{1-}\right\rangle$ remains unpopulated. Once the coupling strength exceeds $0.27\omega_c$, the state $\left\vert \psi_{1-} \right\rangle$ becomes populated, signaling the onset of the cascade decay transition $\left|\psi_{1+}\right\rangle \to \left|\psi_{1-}\right\rangle \to \left|\psi_{0}\right\rangle$. As the coupling strength increases further, the population of $\left|\psi_{1-}\right\rangle$ grows, which corresponds to the rising transition rate $\Gamma_{1+,1-}$ shown in Fig.~\ref{Fig3}(b).
	
	Figures~\ref{Fig3}(e) and ~\ref{Fig3}(f) show the correlation functions versus the coupling strength $g$. The system is resonantly excited on the $\left\vert \psi_0 \right\rangle \to \left\vert \psi_{1+} \right\rangle$ transition. As shown in Fig.~\ref{Fig3}(e), in the SC regime, both the second-order and third-order equal-time correlation functions satisfy $g^{(2)}(0) < 1$ and $g^{(3)}(0) < 1 $. This behavior signifies the emergence of 1PB. In contrast, Fig.~\ref{Fig3}(f) demonstrates how the second-order and third-order equal-time correlation functions vary with $g$ in the USC regime. As shown in Fig.~\ref{Fig3}(f), when $g$ is less than $0.27\omega_c$, although both $g^{(2)}(0)$ and $g^{(3)}(0)$ increase, they remain below $1$. Within this range, the transition rate $\Gamma_{1+,1-}$ corresponding to Fig.~\ref{Fig3}(b) remains $0$, indicating that the process still corresponds to 1PB, a result consistent with our earlier explanation. Once the coupling strength exceeds $0.27\omega_c$, $g^{(2)}(0)$ increases beyond $1$, whereas $g^{(3)}(0)$ remains below $1$. Referring back to Figs.~\ref{Fig3}(b) and ~\ref{Fig3}(d), the transition rate $\Gamma_{1+,1-}$ now begins to increase, indicating the transition between the dressed states $\left|\psi_{1+}\right\rangle$ and $\left|\psi_{1-}\right\rangle$. Hence, in this USC regime, 2PB can occur under resonant single-polariton excitation. The proposed 2PB is well within the reach of current circuit quantum electrodynamics (cQED) technology. A flux qubit galvanically coupled to a transmission-line resonator is a canonical system for achieving the ultrastrong coupling regime (USC)~\cite{frisk2019ultrastrong, yoshihara2017characteristic, RevModPhys.91.025005}, in which the normalized coupling strength of $g/\omega_c \sim 0.6$ has been experimentally demonstrated~\cite{yoshihara2017superconducting}. The coupling phase $\theta$ is tunable \textit{in situ} via an external magnetic flux bias $\delta\psi$ [Fig. 1(a)], enabling the parity-nonconserving condition ($\theta \neq m\pi/2$) for the 2PB effect~\cite{2010Circuit}. The dissipation rates ($\gamma_a=\gamma_{\sigma^{-}}= 10^{-2} \omega_c$) are typical for state-of-the-art superconducting devices~\cite{reed2012realization, RevModPhys.93.025005, PhysRevLett.109.193602}. Furthermore, the second- and third-order correlation functions $g^{(n)}(0)$ to verify the blockade effects are usually measured by Hanbury Brown\textendash Twiss interferometers~\cite{birnbaum2005photon, hennessy2007quantum, dayan2008photon}, routinely employed in cQED experiments to characterize single- and multiphoton sources~\cite{PhysRevLett.107.053602, PhysRevA.87.023809}. Additional results and discussions on 2PB via ultrastrong light-matter coupling
	are provided in the Supplementary Material~\cite{SM}.


	\emph{Conclusions}---Our work has focused on the two-polariton blockade (2PB) in the USC regime by analyzing the high-order optical correlations of the atom-cavity system. Our study uncovers a distinct phenomenon in which 2PB emerges under a resonant single-polariton excitation, in stark contrast to the behavior observed at lower coupling strengths. Additionally, we propose criteria in Eq.~(\ref{Eq2}) based on high-order delay-time correlations to evaluate the 2PB in USC regime. By determining the transition rates and population distributions among various dressed states, we elucidate the physical mechanisms driving this effect. These findings hold significant potential for multi-polariton manipulation.
	
	\vspace{5mm}
	
	\begin{acknowledgements}
		\emph{Acknowledgements}---H. J. is supported by the National Key R\&D Program (Grant No. 
		2024YFE0102400), the National Natural Science Foundation of China (Grants No. 
		11935006 and No. 12421005), and the Hunan Major Sci-Tech Program (Grant No. 
		2023ZJ1010). R. H. is supported by the RIKEN Special Postdoctoral Researchers 
		(SPDR) program. A. M. is supported by the Polish National Science Centre (NCN) 
		under the Maestro Grant No. DEC-2019/34/A/ST2/00081. F. N. is supported in part 
		by the Japan Science and Technology Agency (JST) [via the CREST Quantum 
		Frontiers program Grant No. JPMJCR24I2, the Quantum Leap Flagship Program 
		(Q-LEAP), and the Moonshot R\&D Grant No. JPMJMS256E]. Y.-L. Z. is supported 
		by the Natural Science Foundation of Hunan Province (Grant No. 2025JJ60018) and 
		the Scientific Research Foundation of Education Bureau of Hunan Province (Grant
		No. 24B0866). The authors thank Xun-Wei Xu from Hunan Normal University and 
		Cheng-Song Zhao from Northeastern University for helpful discussions on the 
		numerical simulation of the master equation.
	\end{acknowledgements}

	\bibliography{USCTPB_Refs}

\end{document}